\documentstyle[aas2pp4]{article}

\received{}
\accepted{}
\journalid{000}{00 June 1998}

\begin{document}

{

\title{Ring structure and warp of NGC~5907\\
	-- Interaction with dwarf galaxies\footnotemark[1]}
\footnotetext[1]{The Very Large Array of the National Radio Astronomy 
    Observatory is a facility of the National Science Foundation operated 
    by a cooperative agreement with Associated Universities, Inc.}

\author{
Zhaohui Shang,\altaffilmark{2,3,4} Zhongyuan Zheng,\altaffilmark{2,4} 
Elias Brinks,\altaffilmark{5} Jiansheng Chen,\altaffilmark{2,4} 
David Burstein,\altaffilmark{6} Hongjun Su,\altaffilmark{2}
Yong-IK Byun,\altaffilmark{7}
Licai Deng,\altaffilmark{2,4} 
Zugan Deng,\altaffilmark{4,8}
Xiaohui Fan,\altaffilmark{2,9}
Zhaoji Jiang,\altaffilmark{2,4}
Yong Li,\altaffilmark{6}
Weipeng Lin,\altaffilmark{2,4}
Feng Ma\altaffilmark{3}
Wei-hsin Sun,\altaffilmark{7}
Beverley Wills,\altaffilmark{3}
Rogier A. Windhorst,\altaffilmark{6} 
Hong Wu,\altaffilmark{2,4}
Xiaoyang Xia,\altaffilmark{4,10}
Wen Xu,\altaffilmark{2,6}
Suijian Xue,\altaffilmark{2,4}
Haojing Yan,\altaffilmark{2,6}
Xu Zhou,\altaffilmark{2,4}
Jin Zhu,\altaffilmark{2,4} 
and Zhenlong Zou,\altaffilmark{2,4} 
}

\altaffiltext{2}{Beijing Astronomical Observatory, Chinese Academy of
    Sciences, Beijing 100080, China}
\altaffiltext{3}{Department of Astronomy, University of Texas at Austin,
    Austin, TX 78712, shang@astro.as.utexas.edu}
\altaffiltext{4} {Beijing Astrophysics Center, Beijing, 100871, China}
\altaffiltext{5}{Departamento de Astronom\'{\i}a, Universidad de
    Guanajuato, Apartado Postal 144, Guanajuato, Gto. 36000, Mexico}
\altaffiltext{6}{Department of Physics and Astronomy, Box 871504, Arizona
    State University, Tempe, AZ  85287--1504}
\altaffiltext{7}{Institute of Astronomy, National Central University,
    Chung-Li, Taiwan}
\altaffiltext{8}{Graduate School, Chinese Academy of Sciences, Beijing
    100080, China}
\altaffiltext{9}{Princeton University Observatory, Princeton, New
    Jersey, 08544}
\altaffiltext{10}{Department of Physics, Tianjin Normal University,
    China}

\begin{abstract} 

The edge-on, nearby spiral galaxy NGC~5907 has long been used as the prototype
of a ``non-interacting'' warped galaxy.  We report here the discovery of two
interactions with companion dwarf galaxies that substantially change this
picture.  First, a faint ring structure is discovered around this galaxy that
is likely due to the tidal disruption of a companion dwarf spheroidal galaxy.
The ring is elliptical in shape with the center of NGC~5907 close to one of the
ring's foci. This suggests the ring material is in orbit around NGC~5907. No
gaseous component to the ring has been detected either with deep H$\alpha$
images or in Very Large Array (VLA) HI 21--cm line maps.  The visible material
in the ring has an integrated luminosity $\le 10^8 \, L_\odot$ and its
brightest part has a color  R--I $\sim 0.9$.  All of these properties are
consistent with the ring being a tidally-disrupted dwarf spheroidal galaxy.
Second, we find that NGC~5907 has a dwarf companion galaxy, PGC~54419,
projected to be only 36.9 kpc from the center of NGC~5907, close in radial
velocity ($\Delta V = 45$ km~s$^{-1}$) to the giant spiral galaxy.  This dwarf
is seen at the tip of the HI warp and in the direction of the warp. Hence,
NGC~5907 can no longer be considered ``non-interacting,'' but is obviously
interacting with its dwarf companions much as the Milky Way interacts with its
dwarf galaxies.  These results, coupled with the finding by others that dwarf
galaxies tend to be found around giant galaxies, suggest that tidal 
interaction with companions, even if containing a mere 1\% of the mass 
of the parent galaxy, might be sufficient to excite the warps 
found in the disks of many large spiral galaxies.
\end{abstract}


\section{Introduction}

NGC~5907 is an edge-on spiral galaxy, large in angular size, with a well-known
warp both in its HI gas (\cite{sancisi76}; see also below) and in its optical
disk (\cite{sasaki87}; \cite{morrison94}; \cite{zheng98}).  As such, it has
been a popular galaxy in which to study the vertical luminosity distribution
in a spiral galaxy, not only in the optical (\cite{morrison94};
\cite{sackett94}; \cite{lequeux96}; \cite{lequeux98}; \cite{zheng98}), but
also in the infrared (\cite{rudy97}). Much of the recent interest on NGC~5907
was stimulated by the announcement by Sackett et al. of the detection of a
faint halo around it, sufficient to supply the halo mass of this galaxy.

We present here deep optical images and new VLA HI maps of NGC~5907 that are 
used to reassess the origin of the warp in this galaxy.  In a separate paper 
(\cite{zheng98}), we will use these same data to study the halo of NGC~5907
in detail.

\section{Observations}

Optical observations of NGC~5907 were made as part of the
Beijing-Arizona-Taipei-Connecticut (BATC) Color Survey of the Sky (cf.\
\cite{fan96}).  This survey uses the Beijing Astronomical Observatory (BAO)
0.6/0.9m Schmidt telescope equipped with a $2048 \times 2048$ Ford CCD and a
custom-designed set of 15 intermediate-band filters to do spectrophotometry
for specific 1 deg$^2$ regions of the northern sky. NGC~5907 is in the center
of one of these regions and, given the great interest in its low surface
brightness properties, we obtained long integrations of it in two BATC
filters:  central wavelength 6660$\rm \AA$, bandwidth 480$\rm \AA$ ($\rm
m_{6660}$) and central wavelength 8020$\rm \AA$, bandwidth 260$\rm \AA$ ($\rm
m_{8020}$).

The CCD subtends $58' \times 58 '$ (1.71$''$/pixel) in
the focal plane of the BAO Schmidt.  Out of 43 hours of actual observation of
NGC~5907 at 6660$\rm\AA$, we select the 84 best images for analysis with a
total exposure time of 26.17 hours (\cite{zheng98}).  The telescope was
centered at slightly different positions for these images for greater
reliability of flat-fielding. Figure~1 shows the inner part of the combined
BATC $\rm m_{6660}$ image with low surface brightness features emphasized. We
also obtained 21 hours of integration with the 8020$\rm \AA$ filter, of which
the best 16.7 hours were used.  
By directly comparing with standard broad-band photometry of NGC~5907
(\cite{morrison94}; \cite{rudy97}), we obtain the transformation
between our two bands and R and I bands (\cite{zheng98}).  Within an
error of 0.5 mag arcsec$^{-2}$, our limiting magnitudes correspond to
28.6 and 26.9 mag arcsec$^{-2}$ in R and I, respectively.

As is evident from Figure~1, a luminous ring
encompasses this galaxy.  The part of this ring closest to NGC~5907 can be
seen faintly in Figure~3 of Morrison et al. (1994). In separate papers
(\cite{zheng98}; \cite{chen98}), we detail the data taking and data reduction
processes that permit us to reliably investigate low surface brightness
features on our images.  For the present paper, we present Figure~1 as
evidence that we can reliably flatfield large portions of our images to very
faint surface brightnesses.

We have also used the McDonald Observatory 0.76m telescope (field of view $46'
\times 46'$ with a $2048 \times 2048$ CCD) to obtain broad band R (4.5 hours)
and 30$\rm\AA$--wide H$\alpha$ (7.5 hours) and [OIII] (2.5 hours) images of
NGC~5907.  The narrow band filters are of sufficient width to cover gas
emission from zero velocity to well past the 667 km~s$^{-1}$ heliocentric
recession velocity of NGC~5907.  The ring clearly appears in the 0.76m R
images (ruling out ghost images on Schmidt plates as a possible source), but
does not appear in either narrow band image (the exposures of which are
too short to see the ring in just the faint continuum), 
confirming previous H$\alpha$ observations (\cite{rand96}).

In order to understand the distribution of neutral hydrogen gas around
NGC~5907, 21~cm HI line observations were obtained with the VLA in its
modified C--array (one antenna each from the middle of the East and West arm
having been moved inward to improve coverage of the inner portion of the {\it
uv--}plane). We employed spectral line mode 4, in which the two independently
tunable passbands are used to cover a total velocity range of around 1000
km~s$^{-1}$, centered on 667 km~s$^{-1}$ and with a velocity resolution of 20
km~s$^{-1}$. To optimize sensitivity, both right and left hand polarized
emission are recorded.

At 21 cm, the field of view of the VLA is about 32$'$, large enough to include
the area of interest. As our aim was to go as deep as possible, naturally
weighted maps were made, leading to a synthesized beam of 18$''$, sufficient
to resolve the ring. The total integration time of 6 hours results in a $1
\sigma$ noise level of 0.23 mJy\,beam$^{-1}$. This translates to a column
density of 9 $\times 10^{19}$ atom cm$^{-2}$ assuming a 3$\sigma$
detection in two adjacent channels. The resulting HI map, superimposed on our
optical image of NGC~5907, is shown in Figure~2.  The observations confirm the
impressive warp reported earlier by Sancisi (1976).  On the other hand,
within the quoted sensitivity limits no HI was found to coincide with the
optical ring.

\section{The Ring}

The possible physical interpretations of this ring are either Galactic in
origin --- ``cirrus,'' planetary nebula or supernova remnant --- or physically
associated with NGC~5907. Examination of the whole BATC field of view (only
1/4 of which is shown in Figure~1) shows no other evidence of ring-like
structures near NGC~5907.  In addition, the Burstein--Heiles
(\cite{burstein84}) predicted reddening for this galaxy is only E(B--V) =
0.003.  These two facts make it unlikely that this ring is an artifact due to
Galactic ``cirrus.'' If the ring has a Galactic origin, its ionized gas should
easily have been detected on our narrow band filter images.  In addition, from
our images we measure the $\rm m_{6660} - m_{8020}$ color to be $0.7 \pm 0.3$
mag arcsec$^{-2}$, of the brightest part of the ring ($\rm 26.8 \pm 0.1 \,
m_{6660}$ arcsec$^{-2}$). 
This corresponds to an R--I color of $0.9 \pm 0.4$ based on
our transformation to broad-band photometry, certainly not the color
of a planetary nebula.
These observations, together with the fact that the
field is at high Galactic latitude ($\rm b = +51^\circ$), make it highly
unlikely that this ring could be a Galactic supernova remnant or a planetary
nebula. While we can not completely rule out the possibility that this is a
planetary nebula, it would have to be relatively close (within 1 kpc), very
old and cold, with little ionized gas.

Alternatively, the ring is physically associated with NGC~5907. To
test this hypothesis we derived the luminosity distribution along the
ring and fitted it with an ellipse (Figure~3).  We find that it is
almost perfectly elliptical in shape. Within the errors, the center
of the galaxy falls on the ring's major axis, and is close to one of
its foci, as should be the case if the material in the ring is in
orbit around NGC~5907.  The separation of $1.4'$ between the focus
and the center of NGC~5907 implies an orbit inclination of
45$^{\circ}$ and intrinsic ring dimensions of $13.4' \times 10.3'$
and eccentricity 0.63.

The major axis size is the same as the optical diameter 
of NGC~5907 itself, 43 kpc at an assumed distance of 11 Mpc (using $\rm H_0 =
75$ km~s$^{-1}$ Mpc$^{-1}$).  Thus, based both on the lack of a plausible
alternative and a plausible orbital interpretation of the ring, our conclusion
is that the ring is physically associated with NGC~5907.

How bright is the ring?  Fan et al. (1996) showed that we can flatfield BATC
images to $<$1\% accuracy (see also \cite{zheng98}), and photometrically
calibrate them to 1--2\% absolute accuracy using flux standard stars
(\cite{oke83}). Our images yield an average sky brightnesses of 21.25 mag
arcsec$^{-2}$ at 6660$\rm\AA$ and 19.91 mag arcsec$^{-2}$ at 8020$\rm\AA$,
respectively (on the $\rm A_B$ standard system of Oke \& Gunn 1983). We get
$\rm m_{\rm 6660} \ge 14.7$ for the ring by taking the average surface
brightness for the part of the ring we can see and assuming a complete ring of
similar surface brightness.

In contrast, by replacing all foreground stars with the mean value of
surrounding background, we find that NGC~5907 itself has $\rm m_{6660}
\approx$ 9.9 to a surface brightness of 27 $\rm m_{6660}$ arcsec$^{-2}$.  From
our VLA observations we obtain a rotation velocity of 205 km~s$^{-1}$ and an
outer HI radius of 7.5$'$ (25 kpc), giving a dynamical mass of $1.9 \times
10^{11}$ M$_\odot$.

Based on these measurements, the luminosity of the ring is $\le
1.2$\% that of the galaxy, or an absolute 6660$\rm \AA$ luminosity of $\le
10^8 \, L_\odot$. While it is evident the ring is not of uniform surface
brightness, this upper limit is comparable to the luminosity estimated for the
Sagittarius dwarf galaxy, recently found to be in the process of tidal
disruption by our own Galaxy (\cite{ibata94}.)  The R--I color of $\sim 0.9$
of the brightest part of the ring suggests that the optical emission is mostly
from late-type stars.

The lack of detection of gas in the ring, either in the VLA 21 cm map or the
H$\alpha$ and O[III] images, suggests that gas is a minor constituent of this
ring.  For example, suppose the pre-tidally-distorted dwarf galaxy originally
had an equal amount of gas associated with it as stars (a reasonable
assumption for low--mass, dIrr galaxies), or $\rm \sim 10^8 \, M_\odot$.  If
we distribute this gas evenly along the ring, we would expect a column density
of $4 \times 10^{19}$ atom cm$^{-2}$, close to our detection limit.  Only
modest clumping would be sufficient for us to detect this gas, which we do not
see (cf. Figure~2).  We thus conclude that the ring consists mostly of stars,
in agreement with its measured R--I color.

\section{PGC 54419 and the Warp}

Our VLA map clearly shows the well-known warped HI disk of NGC~5907 (e.g.,
\cite{sancisi76}) extending far beyond the optical disk. Because of the
absence of any nearby galaxy candidate, NGC 5907 has long been considered a
prime example (\cite{sancisi76}) of a non--interacting galaxy with an HI warp.
However, our VLA map reveals that the object PGC 54419, which was once
considered a background object, has a radial velocity of $712.0 \pm 1.5$ km
s$^{-1}$, or only 45 km~s$^{-1}$ from that of the main galaxy.  It therefore is
a dwarf companion at a projected distance of 36.9 kpc ($\sim 11.5'$) from the
nucleus of NGC~5907. It is located near the northern tip and in the
direction of the warp, at some $3.7'$ (12 kpc) in projection to
the west. PGC 54419 is likely closer to NGC~5907 than the Magellanic Clouds
are to the Milky Way.

We obtain $\rm m_{6660} = 15.34 \pm 0.02$ and $\rm m_{8020} = 15.33 \pm 0.02$
for PGC 54419, with $\rm m_{6660} - m_{8020} = 0.01 \pm 0.03$ (R--I $\sim
0.2$).  From Fan et al. (1996), we estimate its V mag to be essentially the
same as its $\rm m_{6660}$ mag, or V = 15.34.  The angular size of PGC 54419
is $54.6'' \times 31.4''$ ($2.9 \times 1.7$ kpc), yielding an inclination of
$56.6^\circ$ (based on an intrinsic axial ratio of 0.2).  From the VLA data we
observe an HI profile with full width at 20\% maximum of 88 km~s$^{-1}$,
giving an inclination-corrected rotation velocity of 53 km
s$^{-1}$.  From this, we measure the mass of PGC 54419 to be $9.5 \times 10^8$
$\rm M_\odot$ (0.5\% the mass of NGC~5907), and its mass-to-light ratio to be
12 in solar units.  From the VLA HI profile we measure an HI mass of 3.6$\rm
\times 10^7 M_{\sun}$ for this galaxy.  In all of its measured
characteristics, PGC 54419 is a dwarf irregular galaxy (cf. \cite{roberts94};
and is likely misclassified as a spiral galaxy; see \cite{pgc}).

It is of interest to understand why PGC 54419 was not previously detected in
HI by Sancisi (1976). The object measures, in the channel map at 715 km
s$^{-1}$ in which it is most clearly seen, about $30'' \times 12''$ (at a
position angle of $\sim 65^\circ$) and has an integrated flux density of $14
\pm 2$ mJy. Once smoothed to the resolution obtained by Sancisi ($51'' \times
61''$) the source has become unresolved and its flux density corresponds to
2.6 Kelvin, or less than 4$\sigma$ in Sancisi's maps. The signal in
neighbouring channels falls well below 3$\sigma$ which explains why it was not
detected in the Westerbork data.

\section{Discussion}

We propose that the ring around NGC~5907 was once a dwarf spheroidal galaxy in
orbit around NGC~5907, now tidally-disrupted by its interaction with the
larger galaxy.  A nearby (at 36.9 kpc projected distance) dwarf irregular galaxy
is suggestively situated right at the end of the more prominent warp in
NGC~5907 (Figure~2).  At the very least, NGC~5907 is no longer a
``non-interacting'' galaxy with an HI warp.

Low surface brightness, ring--like structures may be common around galaxies,
as few nearby galaxies have been probed as deeply as NGC~5907. Even our own
galaxy is tidally interacting with its entourage of dwarf galaxies (e.g. the
Sagitarrius dwarf; Magellanic stream) and is a likely candidate for such a
ring-like feature in the future (cf. \cite{johnston95}).  Indeed, models by
Johnston et al. (1995) show that such a ring could exist from the tidal
disruption of a dwarf galaxy in much the form we see it around NGC~5907. A
dwarf spheroidal galaxy would have low velocity dispersion (cf.
\cite{bender92}), so its tidal debris would tend to produce a coherent stream.
The stream would tend to bunch up at apogalacticon, which is near where we see
the highest surface brightness in the NGC~5907 ring.  The models of Johnston
et al. indicate that such rings can persist for $\sim 2$ dynamical time
scales, or $\sim 1$ billion years in the case of NGC~5907 (rotation velocity of
205 km~s$^{-1}$ from our HI data).

The situation is not as clear with regards to how the ring and PGC~54419 are
related to the HI warp in NGC~5907.  The ring and dwarf galaxy combined have
$\sim$1\% of the luminosity of NGC~5907, and likely a similar percentage
of its mass.  On the one hand, Weinberg (1995) points out that, in the case of
the LMC and the Galaxy, a dynamically active halo acts to amplify the tidal
warping effect of a small galaxy on the disk of a much larger galaxy, through
distortion of the halo of the larger parent galaxy. On the other hand, the LMC
is 10\% of the mass of the Milky Way, while the ring and dwarf are much
smaller compared to NGC~5907. It remains to be seen whether realistic models
of tidal interactions of outer HI disks with small companion dwarf galaxies
can produce the warps, both HI and optical seen in NGC~5907.

Related to this issue is that detailed investigation of the HI map shows 
that the vertical distribution of HI is not symmetric on both sides near the 
central region of the galaxy.  On the side of the ring there are weak HI 
features just above 3$\sigma$ level, as may result from tidal interaction 
between the ring material and the disk. Deeper VLA maps of this galaxy will 
be needed to confirm this feature.

Finally, it is well-known that dwarf galaxies tend to group around giant
spiral galaxies.  Such is certainly the case for the Milky Way and M31 in the
Local Group (e.g. \cite{grebel97}).  The study of \cite{zaritsky97} shows
the same to be true around most giant spiral galaxies, even ``isolated'' giant
spirals.  While it would be very hard to show that a galaxy did {\it not}
have dwarf companions, given their low surface brightnesses, it is also
equally hard to show that all galaxies have them.  Only by taking a deep look 
at the environments around ``isolated'' spiral galaxies with warped disks
can this issue be resolved. 

\section{Acknowledgments}

We thank McDonald Observatory for supporting part of the observations
and data reductions, David Spergel for a helpful conversation, the
VLA personnel for efficient service, and the anonymous referee for
helpful comments. This research is supported in part by the Chinese
National Natural Science Foundation (CNNSF) and by the U.S. National
Science Foundation (NSF Grant INT-93-01805).

\begin{figure}
\plotone{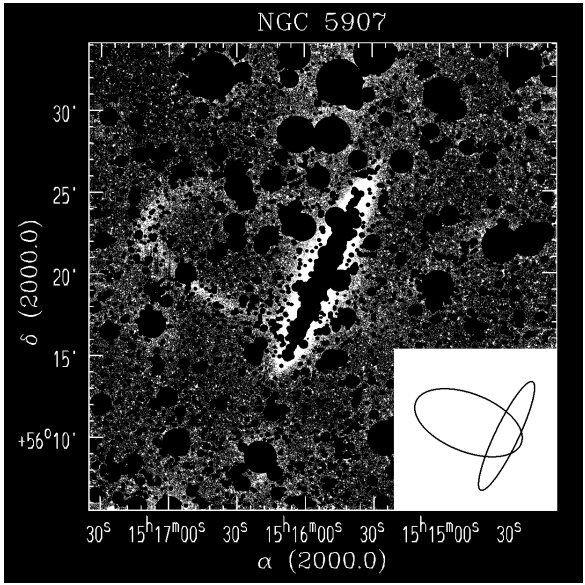}
\caption{This is the combined 26.17 hour BAO Schmidt image of a $28' \times
28'$ subfield of our $58' \times 58'$ image, centered around NGC~5907 taken
through the 6660$\rm \AA$ BATC filter.  All foreground stars, other galaxies
(including PGC 54419) and the dust lane of NGC~5907 have been blanked. This
permits displaying a range of grey levels to enhance the faintest surface
brightness features in this galaxy.  The inset diagram shows the shape of the
ring and its relative orientation with respect to NGC5907.
\label{fig1}}
\end{figure}

\begin{figure}
\plotone{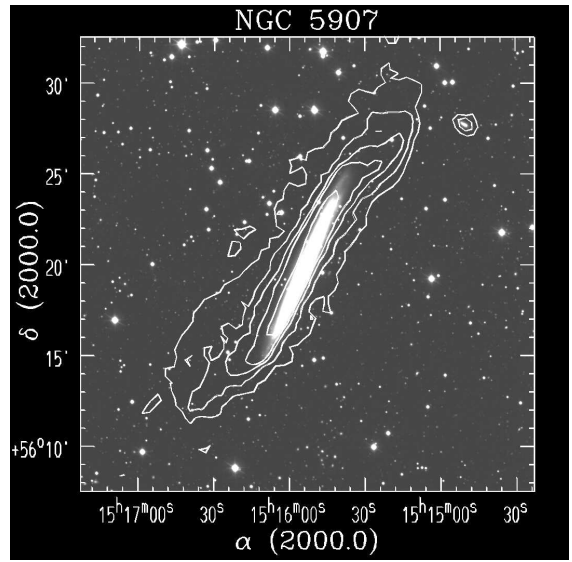}
\caption{The integrated intensity map of 21 cm emission as derived from our 6
hour VLA observation, superimposed on our $\rm m_{6660}$ image, of size $25'
\times 25'$.  The 21 cm contours are 0.69 (3$\sigma$), 4.0, 12.5, 22.5, 75 mJy
beam$^{-1}$ (1 mJy beam$^{-1}$ = 1.85 Kelvin).  The beamwidth is 18$''$.  Note
the discovery of HI at a similar velocity as NGC~5907 in the dwarf galaxy PGC
54419 off to the upper right.
\label{fig2}}
\end{figure}

\begin{figure}
\plotone{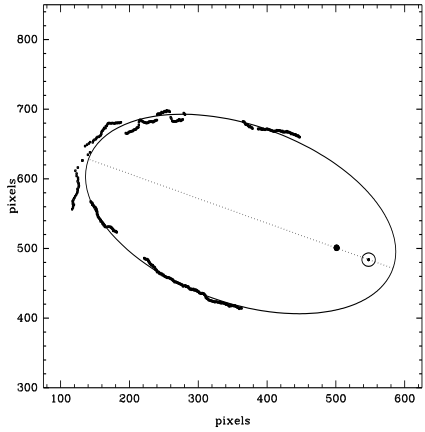}
\caption{The peak brightness distribution along the ring and the
fitted ellipse.  Only the clear part of the ring is measured.  Major
axis is 13.4$'$; minor axis, 7.3$'$ with eccentricity of 0.84. The
small dot is the center of NGC~5907, while the open circle shows the
focus of the ring nearest the galaxy nucleus.  The radius of this
circle (10 pixels) indicates the position error of the focus location.
Also shown is the major axis of the ring.
\label{fig3}}
\end{figure}

}
\end{document}